\documentclass{elsart}
\usepackage{amsmath}
\usepackage{latexsym}
\allowdisplaybreaks    

\newcommand{\seriesref}[1]{Series~\ref{#1}}

\newtheorem{series}{Series}
\newtheorem{series2-6}{Series 2 --}
\newtheorem{series7-15}{Series 7 --}
\newtheorem{series16-21}{Series 16 --}
\newtheorem{series22-27}{Series 22 --}
\newtheorem{series29-37}{Series 29 --}
\newtheorem{series38-47}{Series 38 --}
\newtheorem{series48-52}{Series 48 --}
\begin{document}
\begin{frontmatter}
\title{More Series related to the Euler Series}
\thanks[NFR]{This research has been supported by the Research Council 
of Norway}
\author{Odd Magne Ogreid},
\author{Per Osland}
\address{Department of Physics, University of Bergen, All\'egt.~55,
N-5007 Bergen, Norway}

\begin{abstract}
We present results for infinite series appearing in Feynman diagram 
calculations, many of which are similar to the Euler series.
These include both one-, two- and three-dimensional series.
All these series can be expressed in terms of $\zeta(2)$ and $\zeta(3)$. 
\end{abstract}
\end{frontmatter}

\section{Introduction}
\label{s:intro}

In the evaluation of Feynman integrals, one often needs integrals
and sums related to the dilogarithm and the Riemann zeta function.
This is particularly the case when one considers multi-loop
amplitudes (see \cite{Ritbergen}).
There is actually an intriguing connection between Feynman diagrams, 
topology and number theory, which has recently been elucidated by 
several authors,
in particular by Broadhurst \cite{Broadhurst}, Kreimer \cite{Kreimer}
and collaborators
(see also Groote, K\"orner and Pivovarov \cite{Groote}).

Many results of this kind have been compiled by Devoto and Duke \cite{D&D}
and by K\"olbig et al \cite{Kolbig}, in addition to those of the standard
tables \cite{Prudnikov}.
In an earlier paper \cite{OgreidOsland} (henceforth referred to as Paper 
I\footnote{Often we will
refer to results and identities from our first article on this
subject. Whenever we quote e.g.\ equation (I.13) or
(I.B.2) we are referring to equation (13) or (B.2) in
\cite{OgreidOsland}, respectively.}) 
we presented results for sums required in the evaluation of Feynman 
integrals, related to the Euler series. Several of these series 
involve the digamma or psi function. 
One such example is the series
\begin{eqnarray}
\sum_{n=1}^\infty\frac{1}{n^2}[\gamma+\psi(n)]\nonumber
\end{eqnarray}
which equals $\zeta(3)$ when summed.
Here, we present further results of this kind, many of which are obtained
using known properties of hypergeometric functions.
The sums of these new series are of the form 
\begin{eqnarray}
R_3\zeta(3)+R_2\zeta(2)+R_0\nonumber
\end{eqnarray}
where $R_i$ are all rational numbers.

The starting point of this article will be the well-known result:
\begin{series}\label{series:series1}
\begin{eqnarray}
\sum_{n=1}^\infty\frac{1}{n(n+1)}=1\label{s1}
\end{eqnarray}
\end{series}
This is easily found by recognizing the sum as $\tfrac{1}{2}\ {}_2F_1(1,1;3;1)$
and then using (\ref{2F1unity}).

Using the definition of the Riemann zeta function 
along with partial fractioning, 
we find the following results as immediate corollaries of 
\seriesref{series:series1}:
\begin{series2-6}
\begin{eqnarray}
&&\sum_{n=1}^\infty\frac{1}{n^2(n+1)}=\zeta(2)-1\label{s2}\\
&&\sum_{n=1}^\infty\frac{1}{n(n+1)^2}=-\zeta(2)+2\label{s3}\\
&&\sum_{n=1}^\infty\frac{1}{n^3(n+1)}=\zeta(3)-\zeta(2)+1\label{s4}\\
&&\sum_{n=1}^\infty\frac{1}{n^2(n+1)^2}=2\zeta(2)-3\label{s5}\\
&&\sum_{n=1}^\infty\frac{1}{n(n+1)^3}=-\zeta(3)-\zeta(2)+3\label{s6}
\end{eqnarray}
\end{series2-6}
\addtocounter{series}{5}
The generalization of this type of series is well-known, and is found in
(5.1.24.8) of \cite{Prudnikov1}. These results 
will be frequently used throughout the proofs.  

\section{One-dimensional series}
\label{s:one-d}

We now turn our attention to some one-dimensional series which bear similarity
to the Euler series as well as to those studied in Paper I.

\begin{series7-15}
\begin{eqnarray}
&&\sum_{n=1}^\infty\frac{1}{n(n+1)}[\gamma+\psi(n)]=1\label{s8}\\
&&\sum_{n=1}^\infty\frac{1}{n(n+1)}[\gamma+\psi(1+n)]=\zeta(2)\label{s7}\\
&&\sum_{n=1}^\infty\frac{1}{n(n+1)}[\gamma+\psi(2+n)]=2\label{s9}\\
&&\sum_{n=1}^\infty\frac{1}{n^2(n+1)}[\gamma+\psi(n)]=\zeta(3)-1\label{s14}\\
&&\sum_{n=1}^\infty\frac{1}{n^2(n+1)}[\gamma+\psi(1+n)]=2\zeta(3)-\zeta(2)
\label{s13}\\
&&\sum_{n=1}^\infty\frac{1}{n^2(n+1)}[\gamma+\psi(2+n)]=2\zeta(3)+\zeta(2)-3
\label{s15} \\
&&\sum_{n=1}^\infty\frac{1}{n(n+1)^2}[\gamma+\psi(n)]=-\zeta(3)-\zeta(2)+3
\label{s11}\\
&&\sum_{n=1}^\infty\frac{1}{n(n+1)^2}[\gamma+\psi(1+n)]=-\zeta(3)+\zeta(2)
\label{s10}\\
&&\sum_{n=1}^\infty\frac{1}{n(n+1)^2}[\gamma+\psi(2+n)]=-2\zeta(3)+3
\label{s12}
\end{eqnarray}
\end{series7-15}
\addtocounter{series}{9}

We prove Series \ref{s7}. 
The others follow as corollaries of this result by using the recurrence 
relation 
(\ref{psirecurrence}), partial fractioning, Series \ref{s2}--\ref{s6}, 
(I.B.1) and (I.B.2).
\begin{pf*}{Proof of Series \ref{s7}.}
We start by using the integral representation (\ref{psiintegral}) of the
psi function before summing over $n$:
\begin{eqnarray}
&&\sum_{n=1}^\infty\frac{1}{n(n+1)}[\gamma+\psi(1+n)]=
\sum_{n=1}^\infty\frac{1}{n(n+1)}\int_0^1{\rm
d}t\frac{1-t^n}{1-t}\nonumber\\
&&=\int_0^1{\rm d}t\frac{1}{1-t}\left[1-\frac{t}{2}\ {}_2F_1(1,1;3;t)\right]
\nonumber\\
&&=\int_0^1{\rm d}t\frac{1}{1-t}
\left\{1-\frac{1}{t}\left[t+(1-t)\log(1-t)\right]\right\}
=-\int_0^1{\rm d}t\frac{\log(1-t)}{t}=\zeta(2)
\nonumber
\end{eqnarray}
We have used (7.3.2.150) of \cite{Prudnikov} to rewrite ${}_2F_1$. In
the last step we used (3.6.1) of \cite{D&D}.
\qed\end{pf*}

Similar relations can also be found involving the trigamma function (see
Appendix A.2):
\begin{series16-21}
\begin{eqnarray}
&&\sum_{n=1}^\infty\frac{1}{n}\psi^\prime(n)=2\zeta(3)\label{s18}\\
&&\sum_{n=1}^\infty\frac{1}{n}\psi^\prime(1+n)=\zeta(3)\label{s17}\\
&&\sum_{n=1}^\infty\frac{1}{n}\psi^\prime(2+n)=\zeta(3)+\zeta(2)-2\label{s19}\\
&&\sum_{n=1}^\infty\frac{1}{n(n+1)}\psi^\prime(n)=1\label{s21}\\
&&\sum_{n=1}^\infty\frac{1}{n(n+1)}\psi^\prime(1+n)=-\zeta(3)+\zeta(2)
\label{s20}\\
&&\sum_{n=1}^\infty\frac{1}{n(n+1)}\psi^\prime(2+n)=2\zeta(2)-3\label{s22}
\end{eqnarray}
\end{series16-21}
\addtocounter{series}{6}

We prove Series \ref{s17}. The others follow by using the recurrence relation 
(\ref{trigammarecurrence}) and partial fractioning, together with 
Series \ref{s1}--\ref{s6}.
\begin{pf*}{Proof of Series \ref{s17}.}
We start by using the integral representation
(\ref{trigammaintegral}) of the trigamma function:
\begin{eqnarray}
\sum_{n=1}^\infty\frac{1}{n}\psi^\prime(1+n)
&=&-\sum_{n=1}^\infty\frac{1}{n}\int_0^1{\rm d}t\frac{t^n}{1-t}\log t
=\int_0^1\frac{{\rm d}t}{1-t}\log t\log(1-t)\nonumber\\
&=&\int_0^1\frac{{\rm d}t}{t}\log t\log(1-t)=\zeta(3)
\nonumber
\end{eqnarray}
In the last step we have used (3.6.21) of \cite{D&D}.
\qed\end{pf*}

Next, we consider series which are quadratic or bilinear in psi functions.

\begin{series22-27}
\begin{eqnarray}
&&\sum_{n=1}^\infty\frac{1}{n(n+1)}[\gamma+\psi(n)]^2=\zeta(2)+1\label{s24}\\
&&\sum_{n=1}^\infty\frac{1}{n(n+1)}[\gamma+\psi(n)][\gamma+\psi(1+n)]
=\zeta(3)+\zeta(2)\label{s25}\\
&&\sum_{n=1}^\infty\frac{1}{n(n+1)}[\gamma+\psi(n)][\gamma+\psi(2+n)]
=3\label{s1new}\\
&&\sum_{n=1}^\infty\frac{1}{n(n+1)}[\gamma+\psi(1+n)]^2=3\zeta(3)\label{s23}\\
&&\sum_{n=1}^\infty\frac{1}{n(n+1)}[\gamma+\psi(1+n)][\gamma+\psi(2+n)]
=2\zeta(3)+\zeta(2)\label{s2new}\\
&&\sum_{n=1}^\infty\frac{1}{n(n+1)}[\gamma+\psi(2+n)]^2
=\zeta(2)+3\label{s3new}
\end{eqnarray}
\end{series22-27}
\addtocounter{series}{6}

We prove Series \ref{s23}. The others are immediate corollaries 
that follow from using
(\ref{psirecurrence}) along with some of the results derived earlier in this
chapter.

\begin{pf*}{Proof of Series \ref{s23}.}
We start by using the integral representation (\ref{psiintegral}) of the
psi function:
\begin{eqnarray}
&&\sum_{n=1}^\infty\frac{1}{n(n+1)}[\gamma+\psi(1+n)]^2
=\sum_{n=1}^\infty\frac{1}{n(n+1)}
\int_0^1{\rm d}t\frac{1-t^n}{1-t}\int_0^1{\rm d}s\frac{1-s^n}{1-s}
\nonumber\\
&&=\int_0^1\frac{{\rm d}t}{1-t}\int_0^1\frac{{\rm d}s}{1-s}
\left[1-\frac{t}{2}\ {}_2F_1(1,1;3;t)-\frac{s}{2}
\ {}_2F_1(1,1;3;s)\right.\nonumber\\
&&\left.\phantom{=\int_0^1\frac{{\rm d}t}{1-t}\int_0^1\frac{{\rm d}s}{1-s}[}
+\frac{st}{2}\ {}_2F_1(1,1;3;st)\right]\nonumber\\
&&=\int_0^1\frac{{\rm d}t}{1-t}\int_0^1\frac{{\rm d}s}{1-s}
\left\{1-\frac{1}{t}\left[t+(1-t)\log(1-t)\right]\right.\nonumber\\
&&\phantom{=\int_0^1\frac{{\rm d}t}{1-t}\int_0^1\frac{{\rm d}s}{1-s}\{}
-\frac{1}{s}\left[s+(1-s)\log(1-s)\right]\nonumber\\
&&\left.\phantom{=\int_0^1\frac{{\rm d}t}{1-t}\int_0^1\frac{{\rm d}s}{1-s}\{}
+\frac{1}{st}\left[st+(1-st)\log(1-st)\right]\right\}\nonumber
\end{eqnarray}
Here, we have used (7.3.2.150) of \cite{Prudnikov}. We continue to simplify
this expression:
\begin{eqnarray}
&&\int_0^1\frac{{\rm d}t}{1-t}\int_0^1\frac{{\rm d}s}{1-s}
\left[(1-st)\frac{\log(1-st)}{st}\right.\nonumber\\
&&\left.\phantom{\int_0^1\frac{{\rm d}t}{1-t}\int_0^1\frac{{\rm d}s}{1-s}[}
-(1-t)\frac{\log(1-t)}{t}-(1-s)\frac{\log(1-s)}{s}\right]\nonumber\\
&&=\int_0^1\frac{{\rm d}t}{1-t}\left\{
\int_0^1\frac{{\rm d}s}{1-s}
\left[\frac{\log(1-st)}{st}-\frac{\log(1-t)}{t}\right]\right.\nonumber\\
&&\left.\phantom{=\int_0^1\frac{{\rm d}t}{1-t}\{}
+\int_0^1\frac{{\rm d}s}{1-s}\left[\log(1-t)-\log(1-st)\right]
-\int_0^1{\rm d}s\frac{\log(1-s)}{s}\right\}\nonumber\\
&&=\int_0^1\frac{{\rm d}t}{1-t}\left\{
\frac{1}{t}\int_0^1\frac{{\rm d}s}{1-s}\left[\log(1-st)-\log(1-t)\right]
+\frac{1}{t}\int_0^1\frac{{\rm d}s}{s}\log(1-st)
\right.\nonumber\\
&&\left.\phantom{=\int_0^1\frac{{\rm d}t}{1-t}\{}
-\int_0^1\frac{{\rm d}s}{1-s}\left[\log(1-st)-\log(1-t)\right]
+\zeta(2)\right\}\nonumber
\end{eqnarray}
We combine the first and the third of the integrals inside the curly 
brackets, whereas the second one is evaluated to give 
\begin{eqnarray}
\int_0^1\frac{{\rm d}t}{1-t}\left\{
\frac{1-t}{t}\int_0^1\frac{{\rm d}s}{1-s}\left[\log(1-st)-\log(1-t)\right]
-\frac{1}{t}\mbox{Li}_2(t)+\zeta(2)\right\}.\nonumber
\end{eqnarray}
Next, a change of variables yields: 
\begin{eqnarray}
&&\int_0^1\frac{{\rm d}t}{1-t}\left\{
\frac{1-t}{t}\int_0^1\frac{{\rm d}s}{s}\left[\log(1-t+st)-\log(1-t)\right]
-\frac{1}{t}\mbox{Li}_2(t)+\zeta(2)\right\}\nonumber
\end{eqnarray}
We use (3.14.1) and thereafter (2.2.5) of \cite{D&D}:
\begin{eqnarray}
&&\int_0^1\frac{{\rm d}t}{1-t}\left[
-\frac{1-t}{t}\mbox{Li}_2\left(\frac{-t}{1-t}\right)
-\frac{1}{t}\mbox{Li}_2(t)+\zeta(2)\right]\nonumber\\
&&=\int_0^1\frac{{\rm d}t}{1-t}\left\{
\frac{1-t}{t}\left[\mbox{Li}_2(t)+\frac{1}{2}\log^2(1-t)\right]
-\frac{1}{t}\mbox{Li}_2(t)+\zeta(2)\right\}
\nonumber\\
&&=\int_0^1\frac{{\rm d}t}{1-t}\left[\zeta(2)-\mbox{Li}_2(t)\right]
+\frac{1}{2}\int_0^1\frac{{\rm d}t}{t}\log^2(1-t)
=2\zeta(3)+\zeta(3)=3\zeta(3)\nonumber
\end{eqnarray}
In the last step we have used (3.8.9) and (3.6.9) of \cite{D&D}.
\qed\end{pf*}

\section{Two-dimensional series}
\label{s:two-d}

Next, we present results for the sums of several two-dimensional
series. Many of these are proved by using results from the one-dimensional
series of Section 2 and from Paper I.

\begin{series}
\begin{eqnarray}
\sum_{n=1}^\infty\sum_{k=1}^\infty\frac{1}{nk(n+k)}=2\zeta(3)
\end{eqnarray}
\end{series}
\begin{pf*}{Proof.}
\begin{eqnarray}
\sum_{n=1}^\infty\sum_{k=1}^\infty\frac{1}{nk(n+k)}
=\sum_{k=1}^\infty\frac{1}{k^2}[\gamma+\psi(1+k)]=2\zeta(3)\nonumber
\end{eqnarray}
Here, we have used (I.10) and thereafter (I.B.2).
\qed\end{pf*}

\begin{series29-37}
\begin{eqnarray}
&&\sum_{n=0}^\infty\sum_{k=1}^\infty\frac{1}{k(n+k)(1+n+k)}=\zeta(2)
\label{s27}\\
&&\sum_{n=0}^\infty\sum_{k=1}^\infty\frac{1}{k(n+k)^2(1+n+k)}=
2\zeta(3)-\zeta(2)\label{s30}\\
&&\sum_{n=0}^\infty\sum_{k=1}^\infty\frac{1}{k(n+k)(1+n+k)}[\gamma+\psi(k)]
=\zeta(3)\label{s28}\\
&&\sum_{n=0}^\infty\sum_{k=1}^\infty\frac{1}{k(n+k)(1+n+k)}[\gamma+\psi(1+k)]
=2\zeta(3)\label{s29}\\
&&\sum_{n=0}^\infty\sum_{k=1}^\infty\frac{1}{k(n+k)(1+n+k)}[\gamma+\psi(1+n)]
=2\zeta(3)\label{s31}\\
&&\sum_{n=0}^\infty\sum_{k=1}^\infty\frac{1}{k(n+k)(1+n+k)}[\gamma+\psi(n+k)]
=\zeta(3)+\zeta(2)\label{s34}\\
&&\sum_{n=0}^\infty\sum_{k=1}^\infty\frac{1}{k(n+k)(1+n+k)}[\gamma+\psi(1+n+k)]
=3\zeta(3)\label{s33}\\
&&\sum_{n=0}^\infty\sum_{k=1}^\infty\frac{1}{k(n+k)(1+n+k)}[\gamma+\psi(2+n+k)]
=2\zeta(3)+\zeta(2)\label{s32}\\
&&\sum_{n=0}^\infty\sum_{k=1}^\infty\frac{1}{k(n+k)(1+n+k)}
[\gamma+\psi(1+n+2k)]=\tfrac{7}{2}\zeta(3)\label{s35}
\end{eqnarray}
\end{series29-37}
\addtocounter{series}{9}

We need to prove most of these results in different ways. Series \ref{s34}
follows as an immediate corollary of Series \ref{s30} and \ref{s33} after
using (\ref{psirecurrence}). 

\begin{pf*}{Proof of Series \ref{s27}, \ref{s28} and \ref{s29}.}
\begin{eqnarray}
&&\sum_{n=0}^\infty\sum_{k=1}^\infty\frac{1}{k(n+k)(1+n+k)}f(k)\nonumber\\
&&=\sum_{k=1}^\infty\frac{1}{k^2(k+1)}\ {}_2F_1(1,k;2+k;1)f(k)
=\sum_{k=1}^\infty\frac{1}{k^2}f(k)\nonumber
\end{eqnarray}
Here, we have used (\ref{2F1unity}) to rewrite ${}_2F_1$. 
By replacing $f(k)$ with the 
appropriate expression and using the definition of $\zeta(2)$, (I.B.1) or 
(I.B.2) the proof is complete.
\qed\end{pf*}

\begin{pf*}{Proof of Series \ref{s30}.}
\begin{eqnarray}
&&\sum_{n=0}^\infty\sum_{k=1}^\infty\frac{1}{k(n+k)^2(1+n+k)}\nonumber\\
&&=\sum_{n=0}^\infty\sum_{k=1}^\infty\frac{1}{k(n+k)^2}
-\sum_{n=0}^\infty\sum_{k=1}^\infty\frac{1}{k(n+k)(1+n+k)}\nonumber\\
&&=\sum_{k=1}^\infty\frac{1}{k}\psi^\prime(k)-\zeta(2)
=2\zeta(3)-\zeta(2)\nonumber
\end{eqnarray}
In the last steps we used (\ref{trigammaidentity}) and Series \ref{s18}. 
\qed\end{pf*}

\begin{pf*}{Proof of Series \ref{s31}.}
\begin{eqnarray}
&&\sum_{n=0}^\infty\sum_{k=1}^\infty\frac{1}{k(n+k)(1+n+k)}[\gamma+\psi(1+n)]
\nonumber\\
&&=\sum_{n=0}^\infty\frac{1}{(n+1)(n+2)}\ {}_3F_2(1,1,1+n;2,3+n;1)
[\gamma+\psi(1+n)]
\nonumber\\
&&=\sum_{n=1}^\infty\frac{1}{(n+1)(n+2)}\ {}_3F_2(1,1,1+n;2,3+n;1)
[\gamma+\psi(1+n)]
\nonumber
\end{eqnarray}
Here, we have used the fact that the summand vanishes for $n=0$. Next, we
use (7.4.4.40) of \cite {Prudnikov}. Thereafter we use (6.3.2) of
\cite{Abramowitz} and the recurrence relation for the
psi function:
\begin{eqnarray}
&&\sum_{n=1}^\infty\frac{1}{n(n+1)}[\psi(2+n)-\psi(2)][\gamma+\psi(1+n)]
\nonumber\\
&&=\sum_{n=1}^\infty\frac{1}{n(n+1)}
\left[\gamma+\psi(1+n)+\frac{1}{1+n}-1\right][\gamma+\psi(1+n)]\nonumber\\
&&=\sum_{n=1}^\infty\frac{1}{n(n+1)}[\gamma+\psi(1+n)]^2
+\sum_{n=1}^\infty\frac{1}{n(n+1)^2}[\gamma+\psi(1+n)]\nonumber\\
&&\phantom{=}
-\sum_{n=1}^\infty\frac{1}{n(n+1)}[\gamma+\psi(1+n)]=2\zeta(3)\nonumber
\end{eqnarray}
In the last step we use the results of Series \ref{s7}, \ref{s11} and 
\ref{s23}.
\qed\end{pf*}

\begin{pf*}{Proof of Series \ref{s33}.}
We start by using the recurrence relation for the psi
function.
\begin{eqnarray}
&&\sum_{n=0}^\infty\sum_{k=1}^\infty\frac{1}{k(n+k)(1+n+k)}[\gamma+\psi(1+n+k)]
\nonumber\\
&&=\sum_{n=0}^\infty\sum_{k=1}^\infty\frac{1}{k(n+k)(1+n+k)}
\left[\gamma+\psi(2+n+k)-\frac{1}{1+n+k}\right]\nonumber\\
&&=\sum_{n=0}^\infty\sum_{k=1}^\infty
\frac{1}{k(n+k)(1+n+k)}[\gamma+\psi(2+n+k)]\nonumber\\
&&\phantom{=}
-\sum_{n=0}^\infty\sum_{k=1}^\infty\frac{1}{k(n+k)(1+n+k)^2}\nonumber\\
&&=2\zeta(3)+\zeta(2)
-\sum_{n=0}^\infty\sum_{k=1}^\infty\frac{1}{k(n+k)(1+n+k)}
+\sum_{n=0}^\infty\sum_{k=1}^\infty\frac{1}{k(1+n+k)^2}\nonumber\\
&&=2\zeta(3)+\zeta(2)-\zeta(2)
+\sum_{k=1}^\infty\frac{1}{k}\psi^\prime(1+k)
=2\zeta(3)+\zeta(3)=3\zeta(3)\nonumber
\end{eqnarray}
Here, we have used (\ref{trigammaidentity}), Series \ref{s17} and \ref{s27}.
\qed\end{pf*}

\begin{pf*}{Proof of Series \ref{s32}.}
\begin{eqnarray}
&&\sum_{n=0}^\infty\sum_{k=1}^\infty\frac{1}{k(n+k)(1+n+k)}[\gamma+\psi(2+n+k)]
\nonumber\\
&&=\gamma\sum_{n=0}^\infty\sum_{k=1}^\infty\frac{1}{k(n+k)(1+n+k)}
+\left.\frac{{\rm d}}{{\rm d}x}\right|_{x=0}
\sum_{n=0}^\infty\sum_{k=1}^\infty\frac{\Gamma(n+k)}{k\Gamma(2+n+k-x)}
\nonumber\\
&&=\gamma\sum_{k=1}^\infty\frac{1}{k^2(k+1)}\ {}_2F_1(1,k;2+k;1)\nonumber\\
&&\phantom{=}
+\left.\frac{{\rm d}}{{\rm d}x}\right|_{x=0}
\sum_{k=1}^\infty\frac{\Gamma(k)}{k\Gamma(2+k-x)}\ {}_2F_1(1,k;2+k-x;1)
\nonumber\\
&&=\gamma\sum_{k=1}^\infty\frac{1}{k^2}
+\left.\frac{{\rm d}}{{\rm d}x}\right|_{x=0}
\sum_{k=1}^\infty\frac{\Gamma(k)}{k(1-x)\Gamma(1+k-x)}
\nonumber\\
&&=\gamma\sum_{k=1}^\infty\frac{1}{k^2}
+\sum_{k=1}^\infty\frac{1}{k^2}[1+\psi(1+k)]\nonumber\\
&&=\sum_{k=1}^\infty\frac{1}{k^2}[\gamma+\psi(1+k)]
+\sum_{k=1}^\infty\frac{1}{k^2}
=2\zeta(3)+\zeta(2)\nonumber
\end{eqnarray}
Here, we have used (\ref{2F1unity}) to rewrite ${}_2F_1$. In the last step 
we have also used (I.B.2). 
\qed\end{pf*}

\begin{pf*}{Proof of Series \ref{s35}.}
\begin{eqnarray}
&&\sum_{n=0}^\infty\sum_{k=1}^\infty
\frac{1}{k(n+k)(1+n+k)}[\gamma+\psi(1+n+2k)]\nonumber\\
&&=\sum_{n=0}^\infty\sum_{k=1}^\infty\sum_{j=1}^{n+2k}\frac{1}{jk(n+k)(1+n+k)}
\nonumber\\
&&=\sum_{n=0}^\infty\sum_{k=1}^\infty\sum_{j=1}^{2k}\frac{1}{jk(n+k)(1+n+k)}
+\sum_{n=1}^\infty\sum_{k=1}^\infty\sum_{j=1+2k}^{n+2k}\frac{1}{jk(n+k)(1+n+k)}
\nonumber\\
&&=\sum_{n=0}^\infty\sum_{k=1}^\infty\frac{1}{k(n+k)(1+n+k)}[\gamma+\psi(1+2k)]
\nonumber\\
&&\phantom{=}
+\sum_{n=1}^\infty\sum_{k=1}^\infty\sum_{j=1}^{n}\frac{1}{k(n+k)(1+n+k)(j+2k)}
\nonumber\\
&&=\sum_{k=1}^\infty\frac{1}{k^2(k+1)}\ {}_2F_1(1,k;2+k;1)[\gamma+\psi(1+2k)]
\nonumber\\
&&\phantom{=}
+\sum_{n=0}^\infty\sum_{k=1}^\infty\sum_{j=1}^\infty
\frac{1}{k(n+j+k)(1+n+j+k)(j+2k)}\nonumber\\
&&=\sum_{k=1}^\infty\frac{1}{k^2}[\gamma+\psi(1+2k)]\nonumber\\
&&\phantom{=}
+\sum_{k=1}^\infty\sum_{j=1}^\infty\frac{1}{k(j+k)(1+j+k)(j+2k)}
\ {}_2F_1(1,j+k;2+j+k;1)\nonumber\\
&&=\frac{11}{4}\zeta(3)
+\sum_{k=1}^\infty\sum_{j=1}^\infty\frac{1}{k(j+k)(j+2k)}
=\frac{7}{2}\zeta(3)\nonumber
\end{eqnarray}
Here, we have used (\ref{2F1unity}) to rewrite ${}_2F_1$. We also used (I.13) 
and (I.B.4). 
\qed\end{pf*}

\begin{series38-47}
\begin{eqnarray}
&&\sum_{n=1}^\infty\sum_{k=1}^\infty\frac{1}{n(k+1)(n+k)}=\zeta(2)
\label{s36}\\
&&\sum_{n=1}^\infty\sum_{k=1}^\infty\frac{1}{n(k+1)(n+k)^2}=
3\zeta(3)-2\zeta(2)\label{s40}\\
&&\sum_{n=1}^\infty\sum_{k=1}^\infty\frac{1}{n(k+1)(n+k)}[\gamma+\psi(k)]
=\zeta(3)+\zeta(2)\label{s37}\\
&&\sum_{n=1}^\infty\sum_{k=1}^\infty\frac{1}{n(k+1)(n+k)}[\gamma+\psi(1+k)]
=3\zeta(3)\label{s38}\\
&&\sum_{n=1}^\infty\sum_{k=1}^\infty\frac{1}{n(k+1)(n+k)}[\gamma+\psi(2+k)]
=2\zeta(3)+\zeta(2)\label{s39}\\
&&\sum_{n=1}^\infty\sum_{k=1}^\infty\frac{1}{n(k+1)(n+k)}[\gamma+\psi(n)]
=2\zeta(3)\label{s41}\\
&&\sum_{n=1}^\infty\sum_{k=1}^\infty\frac{1}{n(k+1)(n+k)}[\gamma+\psi(1+n)]
=2\zeta(2)\label{s42}\\
&&\sum_{n=1}^\infty\sum_{k=1}^\infty\frac{1}{n(k+1)(n+k)}[\gamma+\psi(2+n)]
=2\zeta(3)+\tfrac{1}{2}\label{s43}\\
&&\sum_{n=1}^\infty\sum_{k=1}^\infty\frac{1}{n(k+1)(n+k)}[\gamma+\psi(n+k)]
=\zeta(3)+2\zeta(2)\label{s45}\\
&&\sum_{n=1}^\infty\sum_{k=1}^\infty\frac{1}{n(k+1)(n+k)}[\gamma+\psi(1+n+k)]
=4\zeta(3)\label{s44}
\end{eqnarray}
\end{series38-47}
\addtocounter{series}{10}

We use different proofs for most of these series. Series \ref{s45} follows
as an immediate corollary of Series \ref{s40} and \ref{s44} after using
(\ref{psirecurrence}). 

\begin{pf*}{Proof of Series \ref{s36}, \ref{s37}, \ref{s38} and \ref{s39}.}
\begin{eqnarray}
\sum_{n=1}^\infty\sum_{k=1}^\infty\frac{1}{n(k+1)(n+k)}f(k)
&=&\sum_{k=1}^\infty\frac{1}{k(k+1)}[\gamma+\psi(1+k)]f(k) \nonumber
\end{eqnarray}
Here, we have used (I.10). By replacing $f(k)$ by the appropriate expression
and using Series \ref{s7}, \ref{s25}, \ref{s23} or \ref{s2new} in connection 
with (\ref{psirecurrence}), the proof is complete. 
\qed\end{pf*}

\begin{pf*}{Proof of Series \ref{s40}.}
\begin{eqnarray}
&&\sum_{n=1}^\infty\sum_{k=1}^\infty\frac{1}{n(k+1)(n+k)^2}\nonumber\\
&&=\sum_{k=1}^\infty\frac{1}{(k+1)^3}{}_4F_3(1,1,1+k,1+k;2,2+k,2+k;1)
\nonumber
\end{eqnarray}
We rewrite this using (7.5.3.4) of \cite{Prudnikov}:
\begin{eqnarray}
&&\sum_{k=1}^\infty\frac{1}{(k+1)^3}\left\{
\frac{(1+k)^2}{k^2}[\gamma+\psi(1+k)]-\frac{(1+k)^2}{k}\psi^\prime(1+k)\right\}
\nonumber\\
&=&\sum_{k=1}^\infty\frac{1}{k^2(k+1)}[\gamma+\psi(1+k)]
-\sum_{k=1}^\infty\frac{1}{k(k+1)}\psi^\prime(1+k)
=3\zeta(3)-2\zeta(2)\nonumber
\end{eqnarray}
Here, we have made use of Series \ref{s13} and \ref{s20}.
\qed\end{pf*}

\begin{pf*}{Proof of Series \ref{s41}, \ref{s42} and \ref{s43}.}
\begin{eqnarray}
&&\sum_{n=1}^\infty\sum_{k=1}^\infty\frac{1}{n(k+1)(n+k)}f(n)\nonumber\\
&&=\sum_{k=1}^\infty\frac{1}{(k+1)^2}f(1)
+\sum_{n=2}^\infty\sum_{k=1}^\infty\frac{1}{n(k+1)(n+k)}f(n)
\nonumber\\
&&=\left[\zeta(2)-1\right]f(1)
+\sum_{n=2}^\infty\frac{1}{n(n-1)}[\psi(1+n)-\psi(2)]f(n)\nonumber\\
&&=\left[\zeta(2)-1\right]f(1)
+\sum_{n=1}^\infty\frac{1}{n(n+1)}[\psi(2+n)-\psi(2)]f(n+1)\nonumber
\nonumber\\
&&=\left[\zeta(2)-1\right]f(1)
+\sum_{n=1}^\infty\frac{1}{n(n+1)}\left[\gamma+\psi(2+n)\right]f(n+1)
\nonumber\\
&&\phantom{=}-\sum_{n=1}^\infty\frac{1}{n(n+1)}f(n+1)
\nonumber
\end{eqnarray}
Here, we first used (I.10). Thereafter we used (6.3.2) of \cite{Abramowitz}.
By replacing $f(n)$ with the appropriate expression and using results
derived earlier, the proof is complete.
\qed\end{pf*}

\begin{pf*}{Proof of Series \ref{s44}.}
\begin{eqnarray}
&&\sum_{n=1}^\infty\sum_{k=1}^\infty\frac{1}{n(k+1)(n+k)}[\gamma+\psi(1+n+k)]
\nonumber\\
&&=\gamma\sum_{n=1}^\infty\sum_{k=1}^\infty\frac{1}{n(k+1)(n+k)}+
\left.\frac{{\rm d}}{{\rm d}x}\right|_{x=0}
\sum_{n=1}^\infty\sum_{k=1}^\infty\frac{\Gamma(n+k)}{n(k+1)\Gamma(1+n+k-x)}
\nonumber\\
&&=\gamma\sum_{k=1}^\infty\frac{1}{k(k+1)}[\gamma+\psi(1+k)]\nonumber\\
&&+\left.\frac{{\rm d}}{{\rm d}x}\right|_{x=0}
\sum_{k=1}^\infty
\frac{\Gamma(1+k)}{(k+1)\Gamma(2+k-x)}\ {}_3F_2(1,1,1+k;2,2+k-x;1)
\nonumber
\end{eqnarray}
We have made use of (I.10). Next, we make use of (7.4.4.40) of 
\cite{Prudnikov} , and obtain
\begin{eqnarray}
&&\gamma\sum_{k=1}^\infty\frac{1}{k(k+1)}[\gamma+\psi(1+k)]\nonumber\\
&&\phantom{=}
+\left.\frac{{\rm d}}{{\rm d}x}\right|_{x=0}
\sum_{k=1}^\infty\frac{\Gamma(k)}{(k+1)\Gamma(1+k-x)}[\psi(1+k-x)-\psi(1-x)]
\nonumber\\
&&=\gamma\sum_{k=1}^\infty\frac{1}{k(k+1)}[\gamma+\psi(1+k)]\nonumber\\
&&\phantom{=}
+\sum_{k=1}^\infty\frac{1}{k(k+1)}
\left\{\psi(1+k)[\gamma+\psi(1+k)]+\zeta(2)-\psi^\prime(1+k)\right\}
\nonumber\\
&&=\sum_{k=1}^\infty\frac{1}{k(k+1)}[\gamma+\psi(1+k)]^2\nonumber\\
&&\phantom{=}
+\zeta(2)\sum_{k=1}^\infty\frac{1}{k(k+1)}
-\sum_{k=1}^\infty\frac{1}{k(k+1)}\psi^\prime(1+k)=4\zeta(3).
\nonumber
\end{eqnarray}
In the final steps we made use of Series \ref{s1}, \ref{s20} 
and \ref{s23}.
\qed\end{pf*}

\begin{series48-52}
\begin{eqnarray}
&&\sum_{n=1}^\infty\sum_{k=0}^\infty\frac{1}{n(k+1)(n+k)}=2\zeta(2)\\
&&\sum_{n=1}^\infty\sum_{k=0}^\infty\frac{1}{n(k+1)(n+k)}[\gamma+\psi(1+k)]
=3\zeta(3)\\
&&\sum_{n=1}^\infty\sum_{k=0}^\infty\frac{1}{n(k+1)(n+k)}[\gamma+\psi(n)]
=3\zeta(3)\\
&&\sum_{n=1}^\infty\sum_{k=0}^\infty\frac{1}{n(k+1)(n+k)}[\gamma+\psi(1+n+k)]
=6\zeta(3)\\
&&\sum_{n=1}^\infty\sum_{k=0}^\infty\frac{1}{n(k+1)(n+k)}[\gamma+\psi(n+k)]
=2\zeta(3)+2\zeta(2)
\end{eqnarray}
\end{series48-52}
\addtocounter{series}{5}
These results all follow immediately from Series \ref{s36}, \ref{s38}, 
\ref{s41}, \ref{s45}, \ref{s44}, (I.B.1) and (I.B.2).

\begin{series}
\begin{eqnarray}
\sum_{n=1}^\infty\sum_{k=0}^\infty\frac{1}{n(n+1)(1+n+k)^2}
=-\zeta(3)+\zeta(2)
\end{eqnarray}
\end{series}
\begin{pf*}{Proof.}
\begin{eqnarray}
\sum_{n=1}^\infty\sum_{k=0}^\infty\frac{1}{n(n+1)(1+n+k)^2}
&=&\sum_{n=1}^\infty\frac{1}{n(n+1)}\psi^\prime(1+n)=-\zeta(3)+\zeta(2)
\nonumber
\end{eqnarray}
We have used (\ref{trigammaidentity}) and Series \ref{s20} in the last steps.
\qed\end{pf*}

\begin{series}
\begin{eqnarray}
\sum_{n=1}^\infty\sum_{k=1}^\infty\frac{1}{k}
\frac{\Gamma(n)\Gamma(k)}{\Gamma(1+n+k)}
=\zeta(3)
\end{eqnarray}
\end{series}
\begin{pf*}{Proof.}
\begin{eqnarray}
&&\sum_{n=1}^\infty\sum_{k=1}^\infty\frac{1}{k}
\frac{\Gamma(n)\Gamma(k)}{\Gamma(1+n+k)}
=\sum_{k=1}^\infty\frac{1}{k^2(k+1)}\ {}_2F_1(1,1;2+k;1)\nonumber\\
&&=\sum_{k=1}^\infty\frac{1}{k^3}=\zeta(3)\nonumber
\end{eqnarray}
Here, we have used (\ref{2F1unity}) to rewrite ${}_2F_1$.
\qed\end{pf*}

\section{A three-dimensional series}
\label{s:three-d}

\begin{series}
\begin{eqnarray}
\sum_{l=0}^\infty\sum_{n=1}^\infty\sum_{k=1}^\infty\frac{1}{nk(1+l+k)(l+n+k)}
=3\zeta(3)
\end{eqnarray}
\end{series}
\begin{pf*}{Proof.}
\begin{eqnarray}
&&\sum_{n=0}^\infty\sum_{k=1}^\infty\sum_{l=1}^\infty\frac{1}{lk(1+n+k)(n+k+l)}
\nonumber\\
&&=\sum_{n=0}^\infty\sum_{k=1}^\infty\frac{1}{k(n+k)(1+n+k)}
[\gamma+\psi(1+n+k)]=3\zeta(3)\nonumber
\end{eqnarray}
Here, we have used (I.10) and Series~\ref{s33}.
\qed\end{pf*}

\appendix
\section{Useful properties of some special functions}
\label{a:trigamma}
\subsection{Three often used identities}
During the proofs we will often use the fact that 
\begin{eqnarray}
{}_2F_1(a,b;c;1)=\frac{\Gamma(c)\Gamma(c-a-b)}{\Gamma(c-a)\Gamma(c-b)},
\label{2F1unity}
\end{eqnarray}
the integral representation for the psi function
\begin{eqnarray}
\gamma+\psi(z)=\int_0^1{\rm d}t\ \frac{1-t^{z-1}}{1-t},\label{psiintegral}
\end{eqnarray}
and the recurrence relation 
\begin{eqnarray}
\psi(1+z)=\psi(z)+\frac{1}{z}\label{psirecurrence}.
\end{eqnarray}
These three results can be found in (7.3.5.2) of \cite{Prudnikov}, (6.3.22)
and (6.3.5)
of \cite{Abramowitz}, respectively.

\subsection{The trigamma function}
The trigamma function is the derivative of the psi function
(cf. eq. (6.4.1) of \cite{Abramowitz}). By taking the
derivative of (\ref{psiintegral}) with respect to $z$
we find the following integral representation for the trigamma function,
\begin{eqnarray}
\psi^\prime(z)=-\int_0^1{\rm d}t\frac{t^{z-1}}{1-t}\log t.
\label{trigammaintegral}
\end{eqnarray}
In the case of unit argument we get (cf. (6.4.2) of \cite{Abramowitz})
\begin{eqnarray}
\psi^\prime(1)=\frac{\pi^2}{6}.
\end{eqnarray}
This function also satisfies the recurrence relation (cf. (6.4.6) of
\cite{Abramowitz}) 
\begin{eqnarray}
\psi^\prime(1+z)=\psi^\prime(z)-\frac{1}{z^2}\label{trigammarecurrence}.
\end{eqnarray}

One way the trigamma function enters into our series is from the following 
type of series:
\begin{eqnarray}
\sum_{n=0}^\infty\frac{1}{(n+a)^2}=
\frac{1}{a^2}\ {}_3F_2(1,a,a;1+a,1+a;1)=\psi^\prime(a)
\label{trigammaidentity}
\end{eqnarray}
where (7.4.4.34) of \cite{Prudnikov} is used in the last step.

\section{Generalizations}
As mentioned in the introduction, all the series presented in this article 
are of the form 
\begin{eqnarray}
R_3\zeta(3)+R_2\zeta(2)+R_0,\label{class}
\end{eqnarray}
where $R_i$ are rational numbers. We may thus define a class of series
which when summed will be of the form (\ref{class}). All the series
presented in this article belong to this particular class of series.
In Paper I we also presented the results 
\begin{eqnarray}
&&\sum_{n=1}^\infty\frac{1}{n^2}[\gamma+\psi(2n)=\tfrac{9}{4}\zeta(3)\\
&&\sum_{n=1}^\infty\frac{1}{n^2}[\gamma+\psi(1+2n)=\tfrac{11}{4}\zeta(3)
\end{eqnarray}
in (I.B.3) and (I.B.4). These could be obtained from (I.B.1) and (I.B.2) 
by replacing $\psi(n)$ and $\psi(1+n)$ with $\psi(2n)$ and $\psi(1+2n)$, 
respectively.

If we make the same
replacement in Series \ref{s8}, \ref{s7}, \ref{s14}, \ref{s13}, \ref{s11}
and \ref{s10}, we may sum the series using methods similar to those 
used in Section 2. We present the results here without proof:
\begin{eqnarray}
&&\sum_{n=1}^\infty\frac{1}{n(n+1)}[\gamma+\psi(2n)]=2\log 2+\tfrac{1}{2}\\
&&\sum_{n=1}^\infty\frac{1}{n(n+1)}[\gamma+\psi(1+2n)]
=\tfrac{1}{2}\zeta(2)+2\log 2\\
&&\sum_{n=1}^\infty\frac{1}{n^2(n+1)}[\gamma+\psi(2n)]
=\tfrac{9}{4}\zeta(3)-2\log 2-\tfrac{1}{2}\\
&&\sum_{n=1}^\infty\frac{1}{n^2(n+1)}[\gamma+\psi(1+2n)]
=\tfrac{11}{4}\zeta(3)-\tfrac{1}{2}\zeta(2)-2\log 2\\
&&\sum_{n=1}^\infty\frac{1}{n(n+1)^2}[\gamma+\psi(2n)]
=-\tfrac{9}{4}\zeta(3)-\tfrac{3}{2}\zeta(2)+6\log 2+\tfrac{3}{2}\\
&&\sum_{n=1}^\infty\frac{1}{n(n+1)^2}[\gamma+\psi(1+2n)]
=-\tfrac{9}{4}\zeta(3)-\tfrac{1}{2}\zeta(2)+6\log 2
\end{eqnarray}

These series contain terms with $\log 2$, and are of the form
\begin{eqnarray}
R_3\zeta(3)+R_2\zeta(2)+R_1\log 2+R_0,\label{extclass}
\end{eqnarray}
where $R_i$ are rational numbers. These series do not belong to the class
(\ref{class}) mentioned above, 
but can be considered members of an extension of this class. 


\end{document}